\documentclass[12pt,a4paper]{article}
\usepackage{amssymb,amsmath}
\usepackage[dvips]{lscape,graphicx}

\newcommand{\ct}{\cite}
\newcommand{\lb}{\label}

\newcommand{\bc}{\begin{center}}
\newcommand{\ec}{\end{center}}
\newcommand{\bd}{\begin{displaymath}}
\newcommand{\ed}{\end{displaymath}}
\newcommand{\be}{\begin{equation}}
\newcommand{\ee}{\end{equation}}
\newcommand{\ba}{\begin{array}}
\newcommand{\ea}{\end{array}}
\newcommand{\bt}{\begin{tabular}}
\newcommand{\et}{\end{tabular}}
\newcommand{\un}{\underline}

\newcommand{\bp}{\begin{picture}}
\newcommand{\ep}{\end{picture}}
\newcommand{\bfi}{\begin{figure}}
\newcommand{\efi}{\end{figure}}

\begin{document}

\vspace{2cm}

\title {\bf A New Bound State $6t + 6\bar t$ and the
Fundamental-Weak Scale Hierarchy in the Standard Model}

\vspace{2cm}

\author{{\large C.D.Froggatt}${}^{1}$,{\large ~\un {L.V.Laperashvili}}${}^{2}$,
{\large ~ H.B.Nielsen}${}^{3}$\\[10mm] \itshape{${}^{1}$ Department of
Physics and Astronomy,}\\[3mm] \itshape{Glasgow University,
Glasgow, Scotland.}\\[3mm] \itshape{${}^{2}$ Institute of Theoretical and Experimental Physics,
 Moscow, Russia.}\\[3mm] \itshape{${}^{3}$ The Niels Bohr
Institute, Copenhagen, Denmark.}}

\date{}

\maketitle

\pagenumbering{arabic}

\begin{abstract}

The multiple point principle, according to which
several vacuum states with the same energy density exist, is put
forward as a fine-tuning mechanism predicting the exponentially
huge ratio between the fundamental and weak scales in the Standard
Model (SM). Using renormalisation group equations for the SM, we
obtain the effective potential in the 2-loop approximation and
investigate the existence of its postulated second minimum at the
fundamental scale. A prediction is made of the existence of a new
bound state of 6 top quarks and 6 anti-top quarks, formed due to
Higgs boson exchanges between pairs of quarks/anti-quarks. This
bound state is supposed to condense in a new phase of the SM
vacuum. The existence of three vacuum states (new, weak and
fundamental) solves the hierarchy problem in the SM.

\end{abstract}

\vspace{0.5cm} \footnoterule{\noindent${}^{1}$ E-mail:
c.froggatt@physics.gla.ac.uk\\ ${}^{2}$ E-mail:
laper@heron.itep.ru\\ ${}^{3}$ E-mail: hbech@nbi.dk}

\section{Cosmological Constant and Multiple Point Principle}

In the present talk we suggest a scenario, using only the pure SM,
in which an exponentially huge ratio between the fundamental
(Planck) and electroweak scales results: $
\frac{\mu_{fund}}{\mu_{ew}} \sim e^{40}.$

In such a scenario it is reasonable to assume the existence of a
simple and elegant postulate which helps us to explain the SM
parameters: couplings, masses and mixing angles. In our model such
a postulate is based on a phenomenologically required result in
cosmology: the cosmological constant is zero, or approximately
zero, meaning that the vacuum energy density is very small. {\it A
priori} it is quite possible for a quantum field theory to have
several minima of its effective potential as a function of its
scalar fields. Postulating zero cosmological constant, we are
confronted with a question: is the energy density, or cosmological
constant, equal to zero (or approximately zero) for all possible
vacua or it is zero only for that vacuum in which we live?

This assumption would not be more complicated if we postulate that
all the vacua which might exist in Nature, as minima of the
effective potential, should have approximately zero cosmological
constant. This postulate corresponds to what we call the Multiple
Point Principle (MPP) \ct{1}.

The MPP postulates: {\it there are many vacua with the same energy
density or cosmological constant, and all cosmological constants
are zero, or approximately zero.}

In the present talk we want to use this principle to solve the
hierarchy problem in the SM.

\section{The renormalisation group equation for the effective
potential}

The renormalisation group (RG) improvement of the effective
potential, which is a function of the scalar field $\phi$ obeys
the Callan-Symanzik equation (see Refs.\ct{2}):
\be
     (M\frac{\partial}{\partial M} + \beta_{m^2}\frac{\partial}
     {\partial m^2} + \beta_{\lambda}
     \frac{\partial}{\partial \lambda} + \beta_g \frac{\partial}{\partial g}
      + \gamma\, \phi\frac{\partial}{\partial \phi})
     V_{eff}(\phi) = 0.      \lb{2}
\ee Here M is a renormalisation mass parameter, $\beta_{m^2}$,
$\beta_{\lambda}$, $\beta_g$ are the RG functions for mass, scalar
field self-interaction and gauge couplings, respectively; $\gamma$
is the anomalous dimension, $g_i$ are gauge coupling constants:
$g_i = (g', g, g_3)$ for $U(1)_{Y(hypercharge)}$, $SU(2)$ and
$SU(3)$ groups of the SM.

{\it From now on $h\stackrel{def}{=}g_t$ is the top-quark Yukawa
coupling constant. And we neglect all Yukawa couplings of light
fermions.}

In the loop expansion of the $V_{eff}$:
\be
        V_{eff} = V^{(0)} + \sum_{n=1} V^{(n)},     \lb{3}
\ee we have $V^{(0)}$ as a tree-level potential of the SM.

The breaking $SU(2)_L\times U(1)_Y \to U(1)_{em}$ is achieved in
the SM by the Higgs mechanism, giving masses to the gauge bosons
$W^{\pm}$, $Z$, the Higgs boson and the fermions.

With one Higgs doublet of $SU(2)_L$, we have the following
tree--level Higgs potential: \be
        V^{(0)} = - m^2 \Phi^{+}\Phi + \frac{\lambda}{2}
        (\Phi^{+}\Phi )^2.                  \lb{4}                  
\ee The vacuum expectation value of the Higgs field $\Phi$ is:
\be
              <\Phi> = \frac{1}{\sqrt 2}\left(
             \ba{c}
             0\\
             v
             \ea
             \right),               \lb{5}
\ee where
\be
 v = \sqrt{\frac{2 m^2}{\lambda}}\approx 246\,\,{\mbox{GeV}}.
 \lb{6}
\ee Introducing a four-component real field $\phi$:
\be
      \Phi^{+}\Phi = \frac{1}{2}\phi^2,    \lb{7}
\ee
we have the following tree-level potential:
\be
     V^{(0)} = - \frac{1}{2} m^2 \phi^2 + \frac{1}{8} \lambda
\phi^4.              \lb{8}                                  
\ee The masses of the gauge bosons $W$ and $Z$, a fermion with
flavor $f$ and the physical Higgs boson $H$ are expressed in terms
of the VEV parameter $v$: \be
          M_W^2 = \frac{1}{4} g^2 v^2,\quad\quad
          M_Z^2 = \frac{1}{4} (g^2 + g'^2) v^2,\quad\quad
          m_f = \frac{1}{\sqrt 2} h_f v,\quad\quad
          M_H^2 = \lambda v^2,         \lb{23}
\ee where $h_f$ are the Yukawa couplings with the flavor $f$.

\section{The second minimum of the effective potential in the
2-loop approximation}

In our paper \ct{3} we have calculated the 2--loop effective
potential in the limit:
\be
      \phi^2 >> v^2, \quad\quad \phi^2 >> m^2,     \lb{44}
\ee using the SM renormalisation group equations in the 2-loop
approximation given by Ref.\ct{4}. We have obtained: \be
 V_{eff}(2-loop) = (\frac{\lambda}{8} + At + Bt^2)\phi^4,
             \lb{50}
\ee where $\,t =\log (\mu/M) = \log(\phi/M)\,$ is the evolution
variable, \be
    A = \frac{1}{8}(\beta_{\lambda}^{(1)} +  \beta_{\lambda}^{(2)}) +
    \frac{\lambda}{2}(\gamma^{(1)} + \gamma^{(2)} + (\gamma^{(1)})^2) +
     \frac{1}{8}\gamma^{(1)}\beta_{\lambda}^{(1)},
                                    \lb{51}
\ee and
$$
     B = \frac{1}{4}\gamma^{(1)}(\beta_{\lambda}^{(1)} +
     4\lambda \,\gamma^{(1)}) + \frac{3}{32\pi^2}\lambda\,\beta_{\lambda}^{(1)}
 +$$
\be
      \frac{3}{256\pi^2}\beta_{g'}^{(1)}(g'^3 + g'g^2)
     + \frac{3}{256\pi^2}\beta_{g}^{(1)}(3g^3 + g'^2g) -
     \frac{3}{16\pi^2}\beta_h^{(1)}h^3.                \lb{47}
\ee Assuming the existence of the two minima of the effective
potential in the simple SM, we have taken the cosmological
constants for both vacua equal to zero, in accord with the MPP.

Then we have the following illustrative qualitative picture:

\centerline{\includegraphics[width=\textwidth]{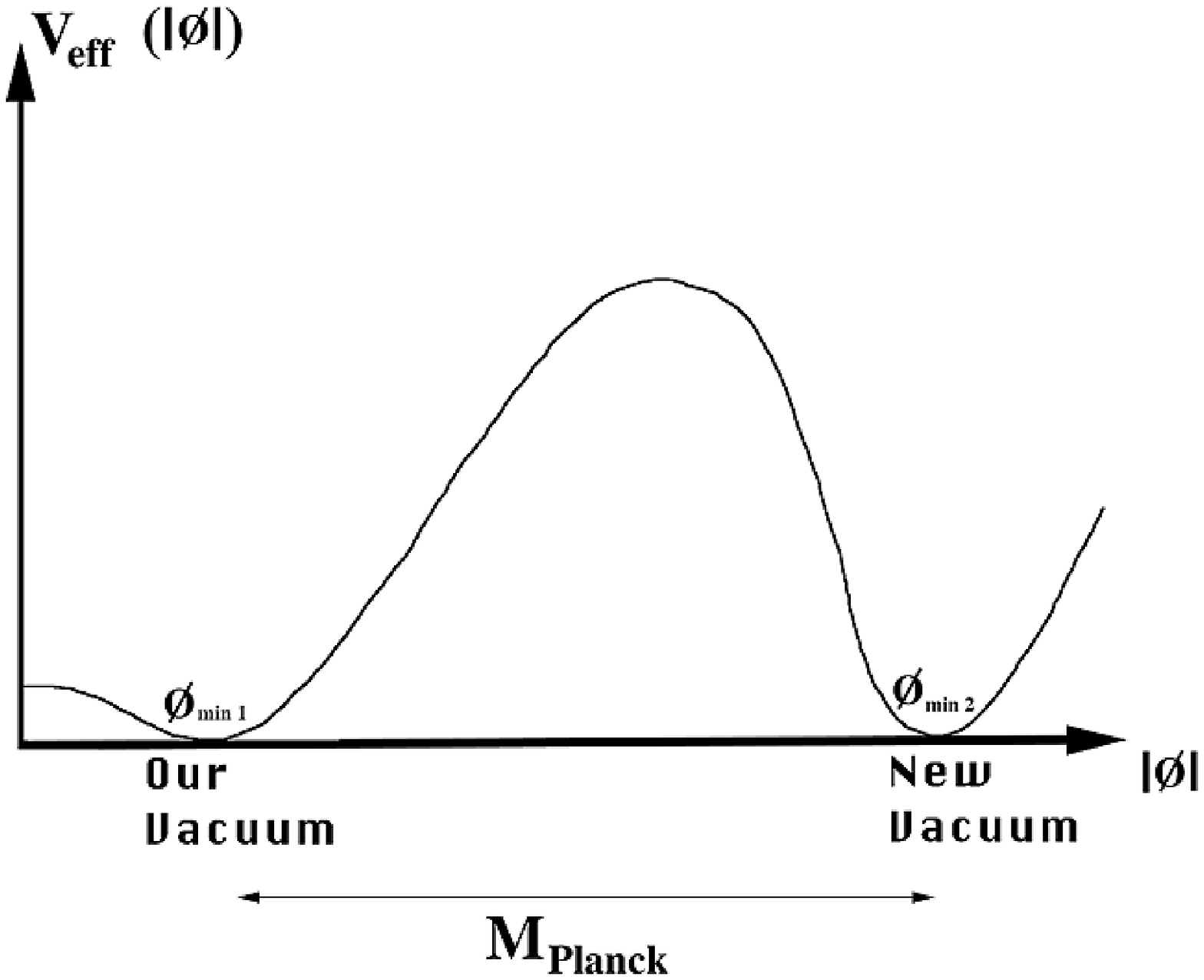}}

\bc
        Fig.1
\ec Here the first minimum: \be
                \phi_{min1} = v = 246 GeV
\ee is the standard "Electroweak scale minimum", in which we live,
and the second one is the non-standard "Fundamental scale
minimum", if it exists.

\section{The Multiple Point Principle requirements}

The MPP requirements for the two degenerate minima in the SM are
given by the following equations:
\be
        V_{eff}(\phi_{min1}) = V_{eff}(\phi_{min2}) = 0,       \lb{53}
\ee
\be
        V'_{eff}(\phi_{min1}) = V'_{eff}(\phi_{min2}) = 0,       \lb{54}
\ee
\be
          V''_{eff}(\phi_{min1}) > 0, \quad\quad
          V''_{eff}(\phi_{min2}) > 0,                        \lb{55}
\ee where
\be
         V'(\phi) = \frac{\partial V}{\partial \phi^2},  \quad
         \quad V''(\phi) = \frac{\partial^2
         V}{\partial(\phi^2)^2}.
                                             \lb{56}
\ee As was shown in Ref.\ct{5}, the degeneracy conditions of MPP
give the following requirements for the existence of the second
minimum in the limit $\phi^2 >> m^2:$
\be
       \lambda_{run}(\phi_{min2}) = 0, \lb{61}
\ee
and
\be
 {\lambda'}_{run}(\phi_{min2}) = 0,    \lb{62}
\ee
what means:
\be
      \beta_{\lambda}(\phi_{min2}, \lambda=0) = 0.  \lb{63}
\ee Using these requirements and the renormalisation group flow
the authors of Ref.\ct{5} computed quite precisely the top quark
(pole) and Higgs boson masses:
\be
M_t = 173\pm 4
\,\,GeV \quad\quad {\mbox{and}} \quad\quad M_H = 135\pm 9\,\, GeV.
                                      \lb{64}
\ee Let us consider now the searching for the fundamental scale
given by these requirements.

\section{The top-quark Yukawa coupling evolution and the second minimum of
the effective potential}

The position of the second minimum of the SM effective potential
essentially depends on the running of gauge couplings and on the
top-quark Yukawa coupling evolution.

Starting from the experimental results \ct{6}, we have:
\be
      M_t = 174.3 \pm 5.1 \,\, {\mbox{GeV}}, \lb{65}
\ee
\be
      M_Z = 91.1872 \pm 0.0021 \,{\mbox{GeV}}, \lb{66}
\ee  and for QCD $\alpha_s$ we have:
\be
       \alpha_3(M_Z)\equiv \alpha_s(M_Z) = 0.117 \pm 0.002.
              \lb{67}
\ee For the running top quark Yukawa coupling constant considered
at the pole mass of t-quark $M_t\,$ the experiment gives:
\be
            h(M_t)\approx 0.95 \pm 0.03.           \lb{70}
\ee Establishing the running of gauge couplings $g',\, g,\,g_3,$
exactly $\, \alpha_Y(t),\,\,\alpha_2(t)\,\,\,$ and $\,\,\,
\alpha_3(t)\,$, in accord with the present experimental data
\ct{6}, and using all experimental results with their
uncertainties we have constructed the evolutions of the inverse
top-quark Yukawa constant: $
      y(t) = \alpha_h^{-1}(t) = 4\pi h^{-2}(t)
$
for different experimental uncertainties (see Fig.2).

\centerline{\includegraphics[width=\textwidth]{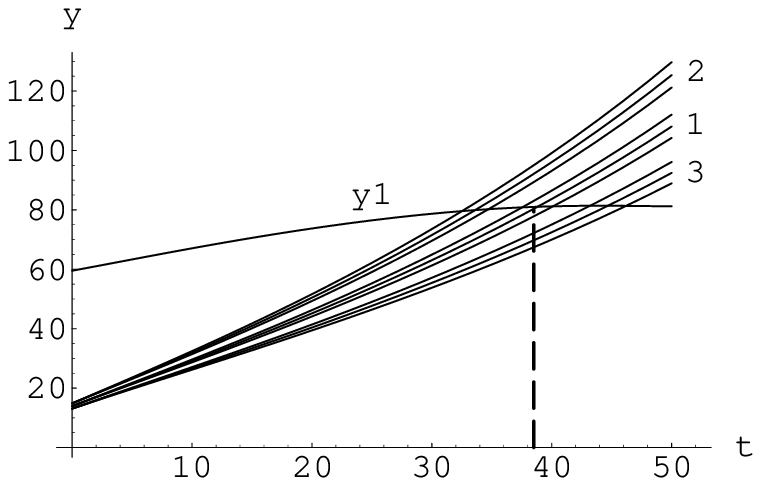}}

\bc  Fig.2 \ec

Three bunches 1(middle), 2(up), 3(down) of curves correspond
respectively to the three values of $h(M_t)=0.95,\,0.92,\,0.98$
given by experiment. The spread of each bunch corresponds to the
experimental values of $\alpha_3(M_Z)=0.117\pm 0.002$. (upper and
lower curves correspond to $\alpha_3(M_Z)=0.115$ and
$\alpha_3(M_Z)=0.119$ respectively).

The curve y1 for $y=\alpha_h^{-1}(t)$ was calculated from the
requirement (\ref{63}): $
        \beta_{\lambda}(\phi_{min2}, \lambda=0) = 0.
$ The intersection of the curve y1 with the evolution of $
\alpha_h^{-1}(t)$ for the experimentally established central
values:
$
      \alpha_s(M_Z) = 0.117\quad {\mbox{and}}\quad
             h(M_t) = 0.95
$
gives us the position of the second minimum of the SM effective
potential at \be  \phi_{min2}\approx 10^{19}\,\,\, GeV. \ee In
general, the experimental uncertainties lead to the following
second minimum position interval: \be \phi_{min2}\approx
10^{16}-10^{22}\,\,\, GeV. \ee Just this position of the second
minimum with given uncertainties predicts the Froggatt-Nielsen's
result \ct{5}:$\,M_H = 135 \pm 9 \,\, GeV$.

The shape of the second minimum at $\mu=10^{19}$ GeV is described
by the curve of Fig.3 where we have used the following
designation:
 \be
   V \stackrel{def}{=}\frac{(16\pi )^4}{24}(\phi_{min2}^{-4})V_{eff},
                                 \lb{85}
\ee

\centerline{\includegraphics[width=\textwidth]{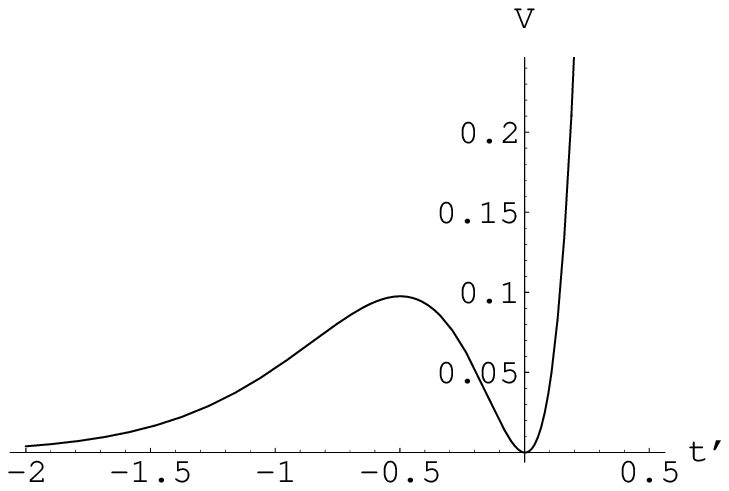}}

\bc Fig.3 \ec

In this scenario the new physics begins at the scale $\sim
10^{19}$ GeV.

\section{A new bound state $6 t + 6 \bar t$, three phases in the
SM and the hierarchy problem}

The MPP is helpful in solving the fine-tuning problems, in
particular, the problem of the electroweak scale being so tiny
compared to the Planck scale.

As is well-known, the quadratic divergencies occur order by order
in the square of the SM Higgs mass, requiring the bare Higgs mass
squared to be fine-tuned again and again as the calculation
proceeds order by order.
If the cut-off reflects new physics entering near the Planck scale
$\Lambda_{Planck}$, then these quadratic divergencies become about
$10^{34}$ times bigger than the final mass squared of the Higgs
particle:
$$
          (\frac{\Lambda_{Planck}}{\Lambda_{electroweak}})^2\sim
          (10^{17})^2 = 10^{34}.
$$
It is clear that an explanation for such a fine-tuning is quite
needed.

Supersymmetry solves the technical hierarchy problem, removing the
divergencies by having a cancellation between fermion and boson
contributions. But the problem of origin of the huge scale ratio
still remains. For example, it exists in the form why the soft
supersymmetry breaking terms are small compared to the fundamental
scale $\Lambda_{Planck}$.

At first sight, it looks difficult to get an explanation of the
cancellation of the quadratic divergencies by fine-tuning, based
on the MPP, which predicts the existence of vacua with degenerate
energy densities. The difficulty is that, from dimensional
arguments, the energy density, or cosmological constant, tends to
become dominated by the very highest frequencies and wave numbers
relevant the Planck scale in our case. In fact, the energy density
has the dimension of energy to the fourth powers, so the modes
with Planck scale frequencies contribute typically
$(10^{17})^4=10^{68}$ times more than those at the electroweak
scale.

Therefore, the only hope of having any sensitivity to electroweak
scale physics is the existence of two degenerate phases in the SM,
which are identical with respect to the modes higher than
electroweak scale frequencies, but deviate by their physics at the
electroweak scale. So, in order to solve the large scale ratio
problem using our MPP we need to have a model with two different
phases that only deviate by the physics at the electroweak scale.

{\it  What could that now be?}

It is obvious that it is necessary to seek a condensation of any
strongly bound states with a binding so strong, in fact, as to
make this bound state tachyonic and to condense it into the
vacuum.

As was shown in papers [7-9], such a bound state can be $ 6 t + 6
\bar t$. Here Higgs scalar particle exchange has an important
special feature. Unlike the exchange of gauge particles, which
lead to alternative signs of the interaction, many top-anti-top
constituents put together lead to attraction in all cases due to
the Higgs scalar boson exchange. This attraction of t and anti-t
quarks by the Higgs exchange is independent of colour.

The bound state of a top quark and an anti-top quark (toponium) is
mainly bound by gluon exchange which is comparable with the Higgs
exchange. But if we now add more top or anti-top quarks, then the
Higgs exchange continues to attract while the gluon exchange
saturates and gets less significant. The maximal binding energy
comes from S-wave $6t+6\bar t$ ground state. The reason is that
the t-quark has 2 spin states and 3 colour states. This means that
by Pauli principle only  6 t-quarks can be put in an S-wave
function, together with 6 anti-t-quarks.
So, in total, we have 6 + 6 = 12 t-constituents together in
relative S-waves.

If we try to put more $t$ and $\bar t$ quarks together, then some
of them will go into a P-wave and the pair binding energy
($E_{binding}$) will decrease by at least a factor of 4.

Calculating the pair binding energy using the Bohr formula for
atomic energy levels (here t and 11t-nucleus), C.D.Froggatt and
H.B.Nielsen [7,8] have obtained the following expression for the
mass squared of the new bound state $6t+6\bar t$:
\be
   m_{bound}^2\approx (12m_t)^2 ( 1 - \frac{33}{8\pi^2}h^4
        +...),
\ee which gives the critical value of h at $m_{bound}^2=0$: $\,
h_{crit}\approx 1.24. $ Taking into account a possible correction
due to the Higgs field quantum fluctuations \ct{9}, we obtained
the following result: \be
                    h_{crit}\approx 1.06\pm 0.18,
\ee what is comparable with the experimental value of the top
Yukawa coupling constant at the electroweak scale:
$h_{exper}(M_t)\equiv g_{t,exper}(M_t)\approx 0.95\pm 0.03.$

\section{The fundamental-electroweak scale hierarchy in the SM}

The requirement of the degeneracy of the three vacua (new,
electroweak and fundamental) solves the hierarchy problem in the
SM.

The central experimental values $h(M_t)=0.95$ and $\alpha_3(M_Z) =
0.117$, together with the vacuum degeneracy conditions
(\ref{61},\ref{63}), predict a second minimum at
$\phi_{min2}\approx 10^{19}$ GeV.

The existence of the second vacuum at $ \phi_{min2}\approx
10^{19}$ GeV  gives a huge ratio between the fundamental and
electroweak scales:
$$
    \frac{\mu_{(fund)}}{\mu_{(ew)}}\sim
    10^{17},
$$
what leads to the prediction of an exponentially huge scale ratio:
$$
\frac{\mu_{(fund)}}{\mu_{(ew)}}\sim e^{40},
$$
in the absence of new physics between the electroweak and
fundamental scales (with the exception of neutrinos).

\section{Acknowledgements}

The speaker (L.V.L.) thanks the financial support by Russian
Foundation for Basic Research, project No.02-02-17379.


\begin{thebibliography}{99}

\bibitem{1}
D.L.Bennett, C.D.Froggatt, H.B.Nielsen, in {\it Proceedings of the
27th International Conference on High Energy Physics, Glasgow,
Scotland, 1994}, Ed. by P.Bussey and I.Knowles (IOP Publishing
Ltd, 1995), p.557; {\it Perspectives in Particle Physics '94}, Ed.
by D.Klabu\u{c}ar, I.Picek and D.Tadi\'{c} (World Scientific,
Singapore, 1995), p.255; ArXiv: hep-ph/9504294;
\newline
D.L.Bennett, H.B.Nielsen, Int.J.Mod.Phys. A {\bf 9}, 5155 (1994).
\bibitem{2}
S. Coleman, E. Weinberg, Phys. Rev. D {\bf 7}, 1888 (1973);
M.Sher, Phys.Rept. {\bf 179}, 274 (1989).
\bibitem{3}
C.D.Froggatt, L.V.Laperashvili, H.B.Nielsen, The fundamental-weak
scale hierarchy in the Standard Model, to be published in Nucl.
Atom. Phys. (Yad. Fiz.); arXiv: hep-ph/0407102.
\bibitem{4}
C.Ford, D.R.T.Jones, P.W.Stephenson, M.B.Einhorn, Nucl.Phys. B
{\bf 395}, 17 (1993).
\bibitem{5}
C.D.Froggatt, H.B.Nielsen, Phys.Lett. B {\bf 368}, 96 (1996).
\bibitem{6}
Particle Data Group, K.Hagiwara et al., Phys.Rev. D {\bf 66},
010001 (2002).
\bibitem{7}
C.D.Froggatt, H.B.Nielsen, Trying to Understand the Standard Model
Parameters, {\it Invited talk by H.B.Nielsen at the XXXI ITEP
Winter School of Physics, Moscow, Russia, 18-26 February 2003};
published in Surveys High Energy Phys. {\bf 18}, 55 (2003); ArXiv:
hep-ph/0308144.
\bibitem{8}
C.D.Froggatt, H.B.Nielsen, Hierarchy Problem and a New Bound
State, in {\it Proc. to the Euroconference on Symmetries Beyond
the Standard Model, Slovenia, Portoroz, 2003}(DMFA, Zaloznistvo,
Ljubljana, 2003), p.73; ArXiv: hep-ph/0312218.
\bibitem{9}
C.D.Froggatt, H.B.Nielsen, L.V.Laperashvili, Hierarchy-Problem and
a Bound State of 6 $t$ and 6 $\bar t$. Invited talk by H.B.Nielsen
at the {\it Coral Gables Conference on Launching of Belle Epoque
in High-Energy Physics and Cosmology (CG2003), 17-21 Dec 2003,
Ft.Lauderdale, Florida, USA} (see Proceedings); ArXiv:
hep-ph/0406110.

\end{thebibliography}
\end{document}